\documentclass[twoside,fleqn, 11pt]{article}
\usepackage{amsmath,amssymb,amsfonts,amsthm,graphics,graphicx,color}
\usepackage[MeX]{polski}
\usepackage[portuguese,spanish,english]{babel}
\usepackage{anysize}
\usepackage{fancyhdr}
\usepackage{easybmat, bm}
\usepackage[T1]{fontenc}
\usepackage[useregional]{datetime2}

\usepackage[small, center, skip=2pt]{caption}
\DeclareCaptionLabelFormat{nospace}{#1#2}

\usepackage{titlesec}
\titleformat{\section}
  {\bfseries  \fontsize{11}{11} \centering} 
  {\thesection.}{0.5em}{}

\titleformat{\subsection}
  {\fontsize{11}{11}\bfseries\sffamily}
  {\thesubsection.}{0.2em}{}
  
\titleformat{\subsubsection}
  {\fontsize{11}{11}\bfseries\sffamily}
  {\subsubsection.}{0.2em}{}  
  
\newenvironment{BLtitle}
    {\begin{center}
    \vspace*{1.2cm}
    \large \bf
    }
    {
    \end{center}
    }  
    
\newenvironment{BLauthor}
    {\begin{center}
    \bf
    }
    {
    \end{center}
    }     
\newenvironment{BLaffiliation}
    {\begin{center}
    \footnotesize
    \vspace{0.2cm}
    }
    {
    \end{center}
  \vspace{0.2cm}
    }

\newenvironment{Summary}
{\begin{minipage}{11.4cm}
\thispagestyle{empty}
\begin{center}
\textsc{Summary\vspace{.2cm}}
\end{center}
\small
}
{
\end{minipage} \\[.2cm]   
}

\newenvironment{Key_words}
{\begin{minipage}{11.4cm}
\thispagestyle{empty}
\small
\textbf{Key words:}
}
{
\end{minipage}    
}

\newenvironment{BLReferences}
{\begin{center}
\textsc{References\vspace{-1.2cm}}
\end{center}

\small
\begin{thebibliography}{xx}
}
{
\end{thebibliography}
}

\usepackage{natbib}
\usepackage{graphicx}

\theoremstyle{plain}
\newtheorem{thm}{Proposition}

\theoremstyle{definition}
\newtheorem{example}{Example}

\newcommand{\h}{\mathcal{H}}
\newcommand{\D}{\mathcal{D}}

\newcommand{\order}{\mathcal{O}}
\DeclareMathOperator{\BIC}{BIC}
\DeclareMathOperator{\BF}{BF}
\DeclareMathOperator{\PBF}{PBF}

\begin{document}
\vspace*{-4.0cm}
\begin{flushright}
\begin{figure}[h!]
\begin{flushleft}
\includegraphics[scale=0.15]{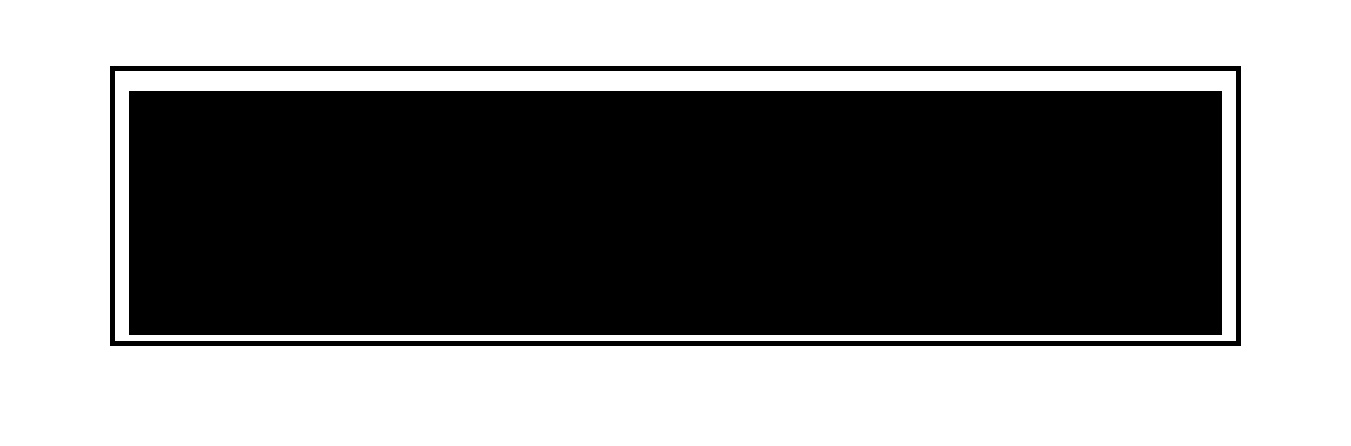}
\end{flushleft}
\end{figure}
\vspace*{-1.2cm}
\footnotesize\textit{\textcolor{black}{DOI: xx.xxxx/xxx-xxx-xxxx } }
\end{flushright}
\begin{center}
\vspace*{0.46cm}
\small \footnotesize
\itshape Article accepted for publication in Biometrical Letters\\[-.01cm]
\today
\end{center} 

\begin{BLtitle}
The Pearson Bayes factor: An analytic formula for computing evidential value from minimal summary statistics
\end{BLtitle}


\begin{BLauthor}
Thomas J. Faulkenberry$^{1}$
\end{BLauthor}


%
\begin{BLaffiliation}
$^1$Department of Psychological Sciences, Tarleton State University, Stephenville, Texas, 76402, USA, e-mail: faulkenberry@tarleton.edu \\
\end{BLaffiliation}

%
%
\begin{Summary}
  In Bayesian hypothesis testing, evidence for a statistical model is quantified by the Bayes factor, which represents the relative likelihood of observed data under that model compared to another competing model. In general, computing Bayes factors is difficult, as computing the marginal likelihood of data under a given model requires integrating over a prior distribution of model parameters. In this paper, I capitalize on a particular choice of prior distribution that allows the Bayes factor to be expressed without integral representation and I develop a simple formula -- the Pearson Bayes factor -- that requires only minimal summary statistics commonly reported in scientific papers, such as the $t$ or $F$ score and the degrees of freedom. In addition to presenting this new result, I provide several examples of its use and report a simulation study validating its performance. Importantly, the Pearson Bayes factor gives applied researchers the ability to compute exact Bayes factors from minimal summary data, and thus easily assess the evidential value of any data for which these summary statistics are provided, even when the original data is not available. 
\end{Summary}
%

\begin{Key_words}
Bayesian statistics; Bayes factor; Pearson Type VI distribution; Summary statistics; $t$-test; analysis of variance.
\end{Key_words}

\setcounter{page}{1}


\section{Introduction}

Across many fields of study, the hypothesis test remains one of the primary tools used by researchers to distinguish signal from noise in empirical observations. Two of the most commonly used tools are the $t$-test and the analysis of variance. Though classically considered within a frequentist paradigm, Bayesian versions of these tests are becoming increasingly popular. Bayesian hypothesis testing makes use of the {\it Bayes factor} \citep{kass1995}, which represents the relative likelihood of observed data under one model -- say, a model hypothesizing a nonzero group difference -- compared to a competing model (e.g., a model hypothesizing a null difference). Bayes factors have many desirable properties for inference, but their difficulty of computation has likely hindered their widespread adoption across the sciences.

One method that has been proposed for computing Bayes factors is the BIC method \citep{kass1995,wagenmakers2007}, which can be even be used when only summary statistics are available \citep{faulkenberry2018,faulkenberry2019}. This popular and well-cited method only requires the use of a pocket calculator, but it suffers from one primary drawback -- it is an {\it approximate} method whose performance suffers in small samples. In this paper, I will present a new formula for computing Bayes factors from minimal summary statistics -- the {\it Pearson Bayes Factor} (PBF). This formula is based on the work of Min Wang and colleagues \citep{wangSun,wang2016} and retains the easy-to-calculate nature of the BIC method while providing the user with an exact Bayes factor computation, regardless of sample size. The formula is summarized in the following proposition:

\begin{thm}[Pearson Bayes factor]\label{mainTheorem}
  Given an analysis of variance summary reported in the form $F(x,y)$, where $x$ equals the between-treatments degrees of freedom and $y$ equals the residual (within-treatments) degrees of freedom, the Bayes factor can be expressed exactly as%
  \[
    {\PBF}_{10} = \frac{\Gamma\Bigl(\frac{x}{2}+\alpha+1\Bigr)\cdot \Gamma\Bigl(\frac{y}{2}\Bigr)}{\Gamma\Bigl(\frac{x+y}{2}\Bigr)\cdot \Gamma(\alpha+1)}\Biggl(\frac{y}{y+xF}\Biggr)^{\alpha-\frac{y}{2}+1}.
    \]
\end{thm}

The paper is roughly organized as follows. After first providing some general statistical background, I will review some of the recent methods used to compute Bayes factors for $t$-tests and analysis of variance designs and illustrate the various problems associated with their use. Then, I will outline the development of the Pearson Bayes factor and provide examples of its computation. Finally, I will present a simulation study which shows that the Pearson Bayes factor compares favorably against other well-known Bayes factors, and in some cases, it has properties which deem it a better choice to use.

\section{Background}
\subsection{$t$-test}
Originally devised by William Sealy Gosset, who published his work under the pseudonym \citet{student1908}, the $t$-test uses the mean and variance of two observed samples to detect differences in the underlying population means. The $t$-test works by first assuming a null hypothesis $\h_0$, and then calculating a $t$-score, which indexes the likelihood of obtaining some sample of observed data under the null hypothesis. If this probability is small, we reject the null hypothesis $\h_0$ in favor of some alternative hypothesis $\h_1$.

Specifically, let us consider the following general setup. Let $x_{ij}$ denote measurements for the $i^{th}$ participant ($i=1,\dots,N$) in the $j^{th}$ experimental condition ($j=1,2$). Further, we assume that the $x_{ij}$ are drawn from independent and normally distributed populations with mean $\mu_j$ and variance $\sigma^2$. Then we can test the hypotheses%
\[
  \h_0:\mu_1=\mu_2\text{  versus  }\h_1:\mu_1\neq \mu_2
\]%
by computing a test statistic%
\[
  t=\frac{\overline{x}_1-\overline{x}_2}{\hat{\sigma}_p/\sqrt{N_{\delta}}}
\]%
whose components are computed as follows. $\overline{x}_j$ represents the sample mean of the measurements in group $j$. $\hat{\sigma}_p$ is the ``pooled'' estimate of $\sigma$, defined by the relationship%
\[
  \hat{\sigma}^2_p = \frac{\hat{\sigma}_1^2(N_1-1)+\hat{\sigma}_2^2(N_2-1)}{N_1+N_2-2},
\]%
where each $\hat{\sigma}_j$ is the sample standard deviation of the measurements in group $j$. Finally, $N_{\delta} = (1/N_1+1/N_2)^{-1}$; this is often called the effective sample size for the experiment. 

Under the null hypothesis $\mathcal{H}_0$, the distribution of these $t$ scores is well known as {\sl Student's $t$ distribution}, with density function
\[
  f_{\nu}(x) = \frac{\Gamma\left(\frac{\nu + 1}{2}\right)}{\sqrt{\nu \pi} \Gamma\left(\frac{\nu}{2}\right)} \left(1+\frac{x^2}{\nu}\right)^{-\frac{\nu+1}{2}},
\]
where $\nu$ represents degrees of freedom of the test and $x\in(-\infty,\infty)$. The cumulative distribution function $F_{\nu}(x)=\int_{-\infty}^x f_{\nu}(u)du$ can then be used to index the probability of observing data at least as extreme as that which we observed under the null hypothesis $\mathcal{H}_0$. Specifically, for an observed $t$-statistic $t_{\text{obs}}$, we compute $P(|x|>t_{\text{obs}}) = 2(1-F_{\nu}(t_{\text{obs}}))$, a quantity commonly known as a $p$-value.  If this probability is small (say, less than 5\%), then one may decide to reject $\mathcal{H}_0$ in favor of $\h_1$ and conclude that $\mu_1\neq \mu_2$, thus implying that the two populations from which we sampled are indeed different.

\subsection{Analysis of variance}
In 1925, Sir Ronald Fisher introduced the analysis of variance \citep{fisher1925}. As a generalization of the $t$-test procedure that works for comparing multiple population means, the analysis of variance (or ANOVA) has become one of the most enduring techniques of hypothesis testing in the experimental sciences. In its simplest form, the analysis of variance is designed to test for differences among multiple group means, making it useful for a wide variety of scientific investigations. Since its introduction, the analysis of variance has become widespread in its popularity and a core topic in most introductory statistics textbooks. In fact, its use is so ubiquitous in the psychological and behavioral sciences that Rouder et al. \citep{rouder2016} referred to the analysis of variance as the ``workhorse'' of experimental psychology.

Roughly, the analysis of variance works by partitioning the total variance in a set of observed data $\D$ into two sources: the variance between experimental treatment groups, and the residual, or left over, variance. Then, one calculates an $F$ statistic, which is defined as the ratio of the between-groups variance to the residual variance. Like the $t$-test above, inference about differences between the treatment groups is then performed by quantifying the likelihood of the observed data $\D$ under a null hypothesis $\h_0$. Specifically, this is done by computing the probability of obtaining the observed $F$ statistic (or greater) under $\h_0$. If this $p$-value is small, this indicates that the data $\D$ are {\it rare} under $\h_0$, so the researcher may reject $\h_0$ in favor of an alternative hypothesis $\h_1$ which posits that there are significant differences among the treatment groups.

\subsection{Alternatives to $p$-values?}
Despite the popularity of the previously-described approaches to testing group differences, there have been many recent criticisms against their use, and more generally, against null hypothesis significance testing \citep[see][]{wagenmakers2007}). In fact, the American Statistical Association has recently recommended against the use of $p$-values and significance testing for scientific inference \citep{asa}. One alternative that has been recommended in its place is the {\it Bayes factor} \citep{jeffreys1961, raftery1995}, which indexes the extent to which the observed data $\D$ are more likely under one hypothesis (e.g., $\h_1$) than another (e.g., $\h_0$). The Bayes factor has several advantages over the $p$-value as a tool for inference, including the ability to index support for either the null hypothesis $\h_0$ or the alternative hypothesis $\h_1$. However, the widespread adoption of Bayes factors has likely been prevented due to the general difficulty of their computation.

In the next section, I will give a conceptual definition of Bayes factors and outline several approaches to computing Bayes factors for $t$-tests and ANOVAs.



\section{Bayes factors}

\subsection{Conceptual definition}

To introduce the Bayes factor, let us first recall Bayes' theorem. Here, we will relax our notation and let $p$ represent both prior probabilities and likelihoods. In later sections, we will denote priors and likelihoods by the more conventional notations $\pi$ and $f$, respectively. Bayes' theorem states%
\begin{equation}\label{eq:bayesRule}
\underbrace{P(\h\mid \D)}_{\substack{\text{Posterior beliefs}\\ \text{about model}}} = \underbrace{P(\h)}_{\substack{\text{Prior beliefs}\\ \text{about model}}} \times \underbrace{\frac{P(\D\mid \h)}{P(\D)}}_{\text{predictive updating factor}}.
\end{equation}%
which tells us that our posterior belief about model $\h$ can be found by taking our \emph{prior} belief about $\h$ and multiplying it by the fraction $P(\D\mid \h)/P(\D)$. This fraction is a \emph{predictive updating factor} in the sense that it indexes how well our data is predicted by $\h$ compared to how well it is predicted by all other possible models. In hypothesis testing, our goal is to directly compare the predictive adequacy of two models $\h_0$ and $\h_1$. To this end, we can use Equation \ref{eq:bayesRule} to derive the following:%
\begin{align*}
  \frac{P(\h_0\mid \D)}{P(\h_1\mid \D)}
  & = \frac{P(\h_0) \cdot \frac{P(\D\mid \h_0)}{P(\D)}}{P(\h_1) \cdot  \frac{P(\D\mid \h_1)}{P(\D)}} 
\end{align*}%
which we can simplify to%
\begin{equation}\label{eq:bayesFactor}
\underbrace{\frac{P(\h_0\mid \D)}{P(\h_1\mid \D)}}_{\substack{\text{posterior beliefs}\\ \text{about models}}} = \underbrace{\frac{P(\h_0)}{P(\h_1)}}_{\substack{\text{prior beliefs}\\ \text{about models}}} \times \underbrace{\frac{P(\D\mid \h_0)}{P(\D\mid \h_1)}}_{\text{predictive updating factor}}.
\end{equation}%
The predictive updating factor
\[
\BF_{01} = \frac{P(\D\mid\h_0)}{P(\D\mid\h_1)}
\]
is called the \emph{Bayes factor}, and as we can see in Equation \eqref{eq:bayesFactor}, it represents the factor by which our relative belief in \(\h_0\) over \(\h_1\) can be updated after observing data. Note that we can easily express our relative belief in $\h_1$ over $\h_0$ by taking reciprocals; that is, $\BF_{10} = 1/\BF_{01}$.

One interesting consequence of Equation \ref{eq:bayesFactor} is that we can use the Bayes factor to compute the posterior probability of $\h_0$ as a function of the prior model probabilities. To see this, consider the following. If we solve Equation \ref{eq:bayesFactor} for the posterior probability $P(\h_0\mid \D)$ and then use Bayes' theorem, we see%
\begin{align*}
  P(\h_0\mid \D) &= \BF_{01} \cdot \frac{P(\h_0)}{P(\h_1)} \cdot P(\h_1\mid \D)\\
                              &= \frac{\BF_{01} \cdot P(\h_0) \cdot P(\D\mid \h_1)\cdot P(\h_1)}{P(\h_1)\cdot P(\D)}\\
  &= \frac{\BF_{01}\cdot P(\h_0)\cdot P(\D\mid \h_1)}{P(\D\mid \h_0)\cdot P(\h_0) + P(\D\mid \h_1)\cdot P(\h_1)}.
\end{align*}%
Dividing both numerator and denominator by the marginal likelihood $P(\D\mid \h_1)$ gives us%
\[
  P(\h_0\mid \D) = \frac{\BF_{01}\cdot P(\h_0)}{\BF_{01}\cdot P(\h_0) + P(\h_1)}.
\]%
Since $\BF_{10} = 1/\BF_{01}$, it can then be shown similarly that
\[
  P(\h_1\mid \D) = \frac{\BF_{10}\cdot P(\h_1)}{\BF_{10}\cdot P(\h_1) + P(\h_0)}.
\]%
In practice, researchers often assume both models are {\it a priori} equally likely, and thus set both $P(\h_0)=P(\h_1) = 0.5$. In this case, we obtain the simplified forms%
\begin{equation}\label{eq:probability}
  P(\h_0\mid \D) = \frac{\BF_{01}}{\BF_{01}+1}, \hspace{1cm} P(\h_1\mid \D) = \frac{\BF_{10}}{\BF_{10}+1}.
\end{equation}

To motivate the work in this paper, let us consider the general problem of computing Bayes factors. From the conceptual definition given in Equation \ref{eq:bayesFactor}, we see that the Bayes factor $\BF_{01}$ is the ratio of likelihoods for $\D$ under $\h_0$ and $\h_1$, respectively. Technically, they are {\it marginal} likelihoods, where the likelihood function $f_i(\D\mid \theta_i)=f(\D\mid \theta_i, \h_i)$ ($i=0,1$) is integrated over the prior distribution $\pi_i(\theta_i)=\pi(\theta_i,\h_i)$ for the vector of model parameters $\theta_i$ associated with model $\h_i$. That is,%
\[
\BF_{01} = \frac{m_0(\D)}{m_1(\D)}
\]%
where%
\[
  m_0(\D) = \int f_0(\D\mid \theta_0)\pi_0(\theta_0)d\theta_0
\]%
and%
\[
  m_1(\D) = \int f_1(\D\mid \theta_1)\pi_1(\theta_1)d\theta_1.
\]

In general, computing these marginal likelihoods can be difficult. One reason is that as sample size increases, the likelihood becomes highly peaked at its maximum. Without prior knowledge of the location of this maximum, numerical integration methods can fail to converge around this highly peaked area. Related to this is the problem of dimensionality: the parameter vector $\theta_i$ is often of high dimension. Combined with the peaked nature of the integrand, numerical methods often have difficulty -- a problem that Thiele, Rouder, and Haaf refer to as ``finding a needle in a haystack'' \citep{thiele2017}. Thus, when possible, it is desirable to make simplifying assumptions which lead to analytic solutions (i.e., without integral representation) for these marginal likelihoods.

\subsection{BIC approximation}
The most popular approach to this problem is the BIC approximation, described in detail by various papers of Kass and Raftery \citep[e.g.,][]{raftery1995, kass1995} and introduced to the applied psychological science community by \citet{wagenmakers2007} and \citet{masson2011}. The basic idea is to construct a second-order Taylor approximation of $\ln m_i(\D) = \ln P(\D\mid \theta_i, \h_i)$ about the posterior mode for $\theta_i$. Of course, by removing all terms beyond the second-derivative, the result is an approximation of the log marginal likelihood. However, \citet{raftery1995} pointed out that with a certain choice of noninformative prior on $\theta_i$ (the {\it unit information prior}), the order of the approximation is $\order(1/\sqrt{n})$. This results in the approximation%
\begin{equation}\label{eq:bic1}
  \ln m_i(\D) = \ln P(\D\mid \hat{\theta}_i,\h_i) - \frac{k_i}{2}\ln n + \order\Bigl(\frac{1}{\sqrt{n}}\Bigr)
\end{equation}%
where $\hat{\theta}_i$ is the maximum likelihood estimate for $\theta_i$ under $\h_i$, $k_i$ is the number of parameters 
in $\h_i$, and $n$ is the total number of observations in $\D$. Equation \ref{eq:bic1} has a fortuitous relation to the Bayesian information criterion (BIC) of \citet{schwarz1978}:%
\begin{equation}\label{eq:bic2}
  \BIC(\h_i)=-2\ln L_i + k_i\ln n,
\end{equation}%
where $L$ is the maximum likelihood estimate for model $\h_i$. Combining Equations \ref{eq:bic1} and \ref{eq:bic2}, we have%
\[
  \ln m_i(\D) = -\frac{1}{2}\BIC(\h_i) + \order\Bigl(\frac{1}{\sqrt{n}}\Bigr),
\]%
or equivalently%
\[
  m_i(\D) \approx \exp\Bigl(-\frac{1}{2}\BIC(\h_i)\Bigr).
\]%
It is then easy to derive the following expression for the BIC approximation to the Bayes factor:
\begin{align}\label{eq:bicBF}
  \BF_{01} = \frac{m_0(\D)}{m_1(\D)} & \approx \frac{\exp\Bigl(-\frac{1}{2}\BIC(\h_0)\Bigr)}{\exp\Bigl(-\frac{1}{2}\BIC(\h_1)\Bigr)}\\ \nonumber
  &= \exp \Biggl(\frac{\BIC(\h_1) - \BIC(\h_0)}{2}\Biggr).
\end{align}

Compared to computing marginal likelihoods by integrating the prior-weighted likelihoods, Equation \ref{eq:bicBF} requires only that we know the BIC values for models $\h_0$ and $\h_1$. For ANOVA models, the calculation is straightforward. In the ANOVA context, the BIC can be calculated \citep{raftery1995} as%
\begin{align*}
BIC &= n\ln (1-R^2)+k\ln n\\
& = n\ln \Biggl(\frac{SSR}{SST}\Biggr) + k\ln n\\
\end{align*}%
where SST represents the total sum of squares for the model (i.e., $\sum_i \sum_j (Y_{ij} - \overline{Y}_{\cdot \cdot})^2)$, SSR represents the residual sum of squares left over after accounting for the model's postulated treatment effects (i.e., $SST - SSA$, where $SSA=r\sum_i(\overline{Y}_{i\cdot} - \overline{Y}_{\cdot \cdot})^2)$, and $k$ equals the number of distinct treatment conditions in the model.

The BIC approach from Equation \ref{eq:bicBF} requires that we have enough ``raw'' data available in order to compute $SSR$ and $SST$. One improvement to Equation \ref{eq:bicBF} is to remove this dependency and make the Bayes factor computation accessible with only summary statistics. \citet{faulkenberry2018} did exactly this, further refining the BIC approximation to produce a formula that can be used by the applied researcher who has access to the commonly reported summary statistics from the ANOVA, usually reported in the standard form $F(x,y)$, where $F$ is the observed $F$-statistic, $x$ is the degrees of freedom between treatments, and $y$ is the residual (within treatments) degrees of freedom. This results in the simple closed form expression%
\begin{equation}\label{eq:BIC_F}
  \BF_{01} \approx \sqrt{n^x\Biggl(1+\frac{Fx}{y}\Biggr)^{-n}}.
\end{equation}%
The advantage of Faulkenberry's formula in Equation \ref{eq:BIC_F} is that researchers may compute approximations to Bayes factors from summary statistics only, whereas the approach of \citet{wagenmakers2007} and \citet{masson2011} that uses the BIC approximation in Equation \ref{eq:bicBF} requires knowing the underlying sum of squares terms in the ANOVA calculation. Thus, Equation \ref{eq:BIC_F} can be useful to researchers who are trying to assess evidential value of observed data from published summary statistics, especially where raw data (or even ANOVA summary tables) are not available \citep{faulkenberry2019ampps}.

The Bayes factor approximation in Equation \ref{eq:BIC_F} can be easily adapted to use with the $t$-test by simply noting that $t^2=F$ and $x=1$. Thus, for a $t$ test with degrees of freedom equal to $\nu$, we have the approximation%
\[
  \BF_{01} \approx \sqrt{n\Biggl(1+\frac{t^2}{\nu}\Biggr)^{-n}}.
\].

Before proceeding further, it will be instructive to consider a simple example of computing the BIC Bayes factor for an ANOVA model.

\begin{example}\label{example1}
Consider the following data representing 6 measurements in each of 3 treatment groups:%
  \begin{center}
  \begin{tabular}{ccc}
    Treatment 1 & Treatment 2 & Treatment 3\\
    \hline
    2 & 5 & 8\\
    3 & 9 & 6\\
    8 & 10 & 12\\
    6 & 13 & 11\\
    5 & 8 & 11\\
    6 & 9 & 12\\
    \hline
  \end{tabular}
\end{center}%
From here, it is a routine exercise to complete the following ANOVA summary table:%
\begin{center}
  \begin{tabular}{ccccc}
    Source & $SS$ & $df$ & $MS$ & $F$\\
    \hline
    Treatments & 84 & 2 & 42 & 7.16\\
    Residual & 88 & 15 & 5.87 & \\
    Total & 172 & 17 & & \\
    \hline
  \end{tabular}
\end{center}%
To use Equation \ref{eq:bicBF}, we need to compute the BIC of $\h_0$ and $\h_1$. First, we compute%
\begin{align*}
  \BIC(\h_1) &= n\ln \Biggl(\frac{SSR}{SST}\Biggr) + k\ln n\\
             &= 18 \ln \Biggl(\frac{88}{172}\Biggr) + 3\ln(18)\\
  &= -3.392.
\end{align*}%
Under $\h_0$, we note that all data are predicted by the grand mean $\mu$ only. Thus, all variation to be considered as residual (i.e., $SSR=172$), and there is $k=1$ treatment condition. This gives us%
\begin{align*}
  \BIC(\h_0) &= 18\ln \Biggl(\frac{172}{172}\Biggr) + 1\ln(18)\\
  &= 2.890.
\end{align*}%
Thus,%
\begin{align*}
  \BF_{01} &\approx \exp \Biggl(\frac{\BIC(\h_1)-\BIC(\h_0)}{2}\Biggr)\\
          &= \exp \Biggl(\frac{-3.392 - 2.890}{2}\Biggr)\\
  &= 0.0432
\end{align*}%
Taking the reciprocal, we see%
\[
  \BF_{10} = \frac{1}{BF_{01}} \approx \frac{1}{0.0432} = 23.15
\]%
indicating that the observed data $\D$ are 23.15 times more likely under $\h_1$ than $\h_0$. Note that this also agrees with the formula using only the summary statistics from the ANOVA table: $F(2,15)=7.16$. Taking from this expression $x=2$ and $y=15$, we evaluate Equation \ref{eq:BIC_F} as%
\begin{align*}
  \BF_{01} &\approx \sqrt{n^x\Biggl(1+\frac{Fx}{y}\Biggr)^{-n}}\\
          &= \sqrt{18^2\Biggl(1 + \frac{7.16\cdot 2}{15}\Biggr)^{-18}}\\
  &= 0.0432,
\end{align*}%
thus demonstrating the equivalence of the approaches from Equations \ref{eq:bicBF} and \ref{eq:BIC_F}.
\end{example}

Though Equations \ref{eq:bicBF} and \ref{eq:BIC_F} provide simple ways to compute Bayes factors from ANOVA summaries, it is important to remember that the BIC formulas represent an {\it approximation} to the Bayes factor. This fact is underscored by the following computation. \citet{sellke2001} showed that under a reasonable class of prior distributions for $p$-values, an upper bound for the Bayes factor can be computed directly from the $p$-value as%
\[
  \BF_{10} \leq -\frac{1}{e\cdot p\ln(p)}.
\]%
Applying this bound to our previous example, we first note that for $F(2,15)=7.16$, the associated $p$-value is $p=0.0066$. This gives us the upper bound%
\begin{align*}
  \BF_{10} & \leq -\frac{1}{e\cdot 0.0066\cdot \ln(0.0066)}\\
          &=11.10.
\end{align*}%
Thus, it is clear that our BIC Bayes factor of 23.15 is quite an overestimate of the actual Bayes factor, so our desire for {\it exact} Bayes factors is quite warranted.

\section{Analytic methods for computing Bayes factors}

Since its introduction by \citet{kass1995}, the BIC Bayes factor has been a popular method for estimating evidential value from empirical data. One of the attractive features of the BIC Bayes factor is that it can be computed with a simple calculator. This makes it easy to implement not only for the researcher, but also in the context of a beginning statistics course. However, it is a large-sample approximation, and the discussion above calls into question its computational stability for smaller examples, especially those which might be used in a traditional statistics course.

Thus, it is desirable to find {\it analytic methods} -- that is, solutions without the need for approximating integrals -- to replace the BIC approximation. In other words, we want a method which removes the integral representation, but unlike the BIC, gives an {\it exact} value for the Bayes factor. In the following sections, I will discuss some historical approaches to developing these methods for the $t$-test and the analysis of variance, followed by a discussion of their limitations.

\subsection{Analytic Bayes factors for the $t$-test}

One approach to developing a Bayesian $t$-test with an analytic solution was provided by \citet{gonen2005}. G\"onen et al. reparameterized $\h_0$ and $\h_1$ in terms of ``effect size'' $\delta = \mu_1-\mu_2$. Their goal was to test%
\[
  \h_0:\delta=0\text{  versus  }\h_1:\delta\neq 0,
\]%
using a parameterization $(\mu=(\mu_1+\mu_2)/2, \delta, \sigma^2)$. Following \citet{jeffreys1961}, they placed a noninformative prior on the mean and variance:%
\[
  \pi(\mu,\sigma^2) \propto 1/\sigma^2
\]%
and assumed, under $\h_1$, that $\delta$ was drawn from a normal distribution centered at 0 with a variance parameter $\sigma^2_a$. This variance term serves as as hyperparameter that needed to be specified in advance by the analyst. Given this setup, \citet{gonen2005} showed%
\begin{equation}
  \label{eq:gonen}
  \BF_{10} = \Biggl(\frac{1+\frac{t^2}{\nu}}{1+\frac{t^2}{\nu(1+n_{\delta}\sigma^2_a)}}\Biggr)^{\frac{\nu+1}{2}} (1+n_{\delta}\sigma^2_a)^{-\frac{1}{2}}.
\end{equation}%
Certainly, Equation \ref{eq:gonen} meets our requirements of (1) being an exact Bayes factor, and (2) being solvable by means of a calculator, but it does have some drawbacks. While the ability for the researcher to specify the prior variance of the effect size under $\h_1$ may be initially appealing, it is fraught with some unfortunate consequences. 

\noindent {\bf Issue 1}: It is a common practice to specify a large value for $\sigma^2_a$, as such a choice would minimize the information provided by the prior and thus potentially better serve as an ``objective'' choice of prior. However, this causes a problem. Consider a fixed data summary; suppose we observed $t=3.00$ with $N_1=N_2=20$. Applying Equation \ref{eq:gonen} with increasing values of $\sigma^2_a$, we see a troubling pattern, shown in the following table:%
\begin{center}
\begin{tabular}{|l|ccccc|}
  \hline
  $\sigma^2_a$ & 1 & 10 & 100 & 1000 & 10000\\
  $\BF_{10}$ & 8.99 & 6.01 & 2.08 & 0.67 & 0.21\\
  \hline
\end{tabular}
\end{center}
Indeed, this table indicates that the G\"onen Bayes factor exhibits {\it Bartlett's paradox}; that is, for fixed information from the data, the Bayes factor tends to 0 (and thus always favors $\h_0$) with increasing prior variance \citep[see also][]{wang2016}.\\

\begin{figure}[h!]
  \centering
  \includegraphics[width=0.8\textwidth]{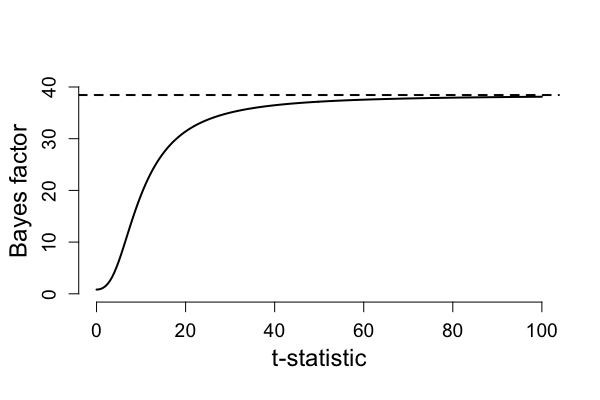}
  \caption{Information paradox: with fixed data under $\h_1$, the G\"onen Bayes factor is bounded with increasing $t$.}
  \label{fig:informationParadox}
\end{figure}

\noindent {\bf Issue 2}: Fix a set of observed data (samples) and suppose these observed data are always generated under $\h_1$. Then as $t$ increases with fixed $\sigma^2_a$, the Bayes factor in Equation \ref{eq:gonen} tends to $(1+n_{\delta}\sigma^2_a)^{\eta/2}$ (see Figure \ref{fig:informationParadox}). However, since the samples are drawn from $\h_1$, we would expect that as $t$ increases without bound, the Bayes factor should also increase without bound. The unexpected asymptote in Figure \ref{fig:informationParadox} is referred to as the {\it information paradox}.

\subsection{Analytic Bayes factors for ANOVA}

Similar to the \citet{gonen2005} $t$-test, it is also possible to calculate analytic solutions for Bayes factors with ANOVA summaries. One approach is to use Zellner's $g$-prior approach, which was originally developed for the context of linear regression models \citep{zellner1986}. In the $g$-prior approach, the standardized regression coefficients $\beta$ are given a multivariate normal prior centered at 0 with a scaled covariance matrix that is proportional to the inverse Fisher information matrix for $\beta$. This scaling factor, denoted $g$, must be chosen by the analyst. Given the construction of the prior, $g$ serves as a scale factor for the analyst to weigh the relative ``importance'' of the prior and the sample in the resulting inference. If $g=1$, then the prior and the sample have the same importance in the inference. If $g=2$, then the prior has half the importance as the sample in the inference. This pattern continues; if $g=n$, then the prior has $1/n^{\text{th}}$ the importance as the sample in the inference.

The $g$-prior Bayes factor can be computed as%
\begin{equation}
  \label{eq:zellner}
  \BF_{10} = (1+g)^{\frac{n-k-1}{2}}\bigl(1+g(1-R^2)\bigr)^{-\frac{n-1}{2}}
\end{equation}%
where $n$ equals the number of observations, $k$ equals the number of predictors in the regression model, and $R^2$ equals the coefficient of determination for the regression model. The analyst must choose the value of $g$, as discussed above. Unfortunately, this leads to the same issues discussed with the \citet{gonen2005} above. First, assuming fixed $n$ and $k$, increasing values of $g$ in Equation \ref{eq:zellner} will cause the Bayes factor to approach 0 (i.e., Bartlett's paradox). Second, increasing values of $R^2$ from data drawn under $\h_1$ will only result in Bayes factors that approach an asymptote, $(1+g)^{(n-k-1)/2}$ (i.e., the information paradox).

In light of these paradoxes (both the G\"onen and Zellner approaches), a common approach for resolution is to place a prior distribution on the relevant scaling parameter (e.g., the $g$ in the Zellner Bayes factor). Two solutions have been recently proposed: (1) the hyper-$g$ approach \citep{liang2008}; and (2) the JZS approach \citep{rouder2012}. Both of these approaches place prior priors on the scale factor $g$ and thus avoid the paradoxes described above. However, the drawback to both of these approaches is that the resulting Bayes factors have integral representation. That is, their computation requires evaluation of at least a one-dimensional integral. With software, this is no problem, but these approaches are inaccessible to beginners (especially students) with no calculus background or programming knowledge.

\subsection{Analytic Bayes factors without integral representation}

Since its introduction by \citet{kass1995}, the BIC Bayes factor has been a popular method for estimating evidential value from empirical data. However, since it is based on a large sample approximation, its utility is limited, particularly in a small sample context. It is especially limited for use in a teaching context, where sample sizes for hand-worked examples tend to be small. Thus, it is desirable to find a calculation method to replace the BIC approximation. Specifically, we want a method which, like the BIC, removes the integral representation, but unlike the BIC, gives an exact value for the Bayes factor.

Recent work by \citet{wangSun} has made possible such calculation. In what follows, we consider our data to be a balanced set of observations $\D = Y_{ij}$ where $i=1,\dots,p$ represents the experimental unit (i.e., the treatment group) and $j=1,\dots,r$ represents the specific replicate within each unit. This gives a total of $n=pr$ observations. On these data, we place a random effects linear model:%
\[
  Y_{ij}=\mu + a_i + \varepsilon_{ij},
\]%
assuming that $a_i\sim \mathcal{N}(0,\sigma^2_a)$ and $\varepsilon \sim \mathcal{N}(0,\sigma^2)$. In this context, the main goal of analysis of variance is to test whether the random effects term $a_i$ is identically 0. We do this by restricting the variability of the random effects term. That is, we want to test%
\[
  \h_0:\sigma^2_a=0 \text{ versus }\h_1:\sigma_a^2\neq 0.
  \]

\begin{figure}[h]
  \centering
  \includegraphics[width=0.8\textwidth]{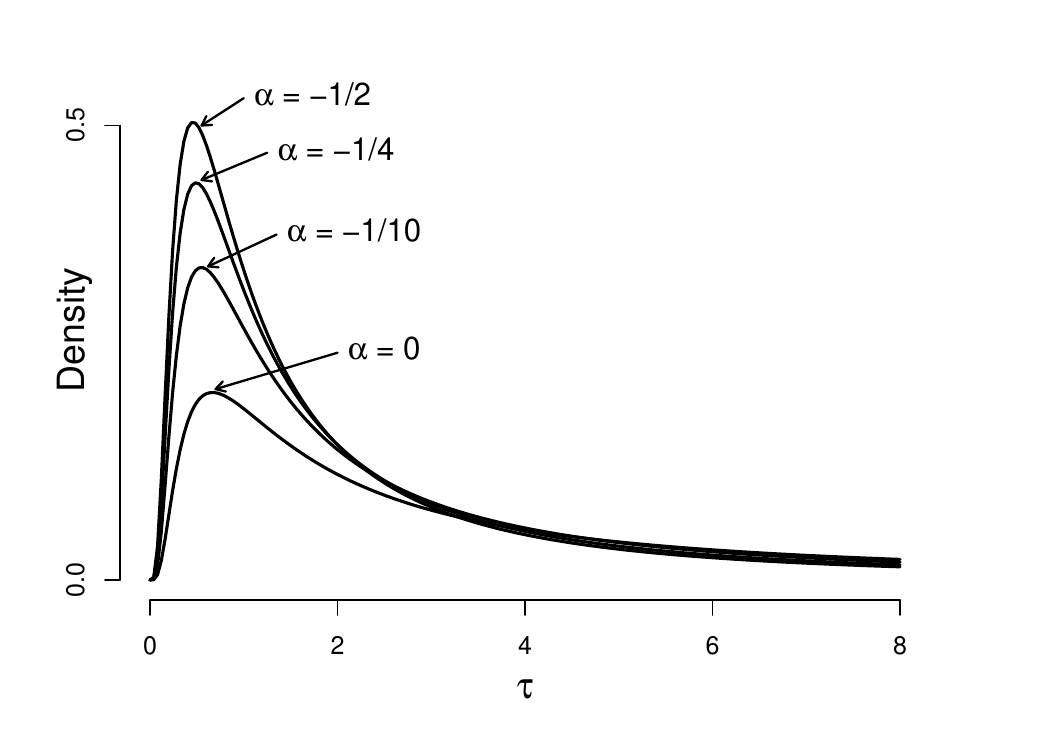}
  \caption{A Pearson Type VI prior for $\tau$, plotted as a function of shape parameter $\alpha$.}
  \label{fig:prior}
\end{figure}

Wang and Sun extended the approach of \citet{gds}, who assumed common, noninformative priors on $\mu$ and $\sigma$ in both $\h_0$ and $\h_1$ and considered a proper prior on the ratio of variance components $\tau = \sigma^2_a/\sigma^2$ under $\h_1$. Under such prior specification, Garcia-Donato and Sun showed%
\begin{equation}
  \label{eq:gds}
  BF_{10} = \int_0^{\infty}(1+\tau r)^{\frac{1-p}{2}}\Biggl(1-\frac{\tau r}{1+\tau r}\cdot \frac{SSA}{SST}\Biggr)^{\frac{1-n}{2}}\cdot \pi(\tau)d\tau
\end{equation}%
To use Equation \ref{eq:gds}, the user must place a prior distribution on the variance components $\tau$. Motivated by the work of \citet{maruyama}, Wang and Sun used a Pearson Type VI distribution to serve as the prior for $\tau$. The Pearson Type VI distribution has three parameters: two shape parameters $\alpha>-1$ and $\beta>-1$ and a scale parameter $\kappa>0$. The density function for the Pearson Type VI prior is given by%
\[
  \pi^{PT}(\tau) = \frac{\kappa(\kappa \tau)^{\beta}(1+\kappa \tau)^{-\alpha-\beta-2}}{\mathcal{B}(\alpha+1, \beta+1)}I_{(0,\infty)}(\tau)
\]%
where $\mathcal{B}(x,y) = \int_0^1t^{x-1}(1-t)^{y-1}dt$ is the standard Beta function. Wang and Sun further restricted $\pi^{PT}$ to one parameter $\alpha \in [-\frac{1}{2}, 0]$ by taking $\kappa=r$ and $\beta = \frac{n-p}{2}-\alpha-2$. A plot of this prior can be seen in Figure \ref{fig:prior}; in this figure, we take the values $r=6$, $p=3$, and $n=18$ from Example \ref{example1}, thus setting $\kappa=6$ and $\beta=\frac{18-3}{2}-\alpha-2$, where $\alpha$ takes specific values $-\frac{1}{2}, -\frac{1}{4}, -\frac{1}{10}, 0$. Further, Figure \ref{fig:prior} shows the effect of varying $\alpha$ on the prior distribution for $\tau$. As we can see, as $\alpha$ increases to 0, $\tau$ becomes more dispersed and less peaked around the mode. This places more prior mass on larger treatment effects than we would see for values of $\alpha$ closer to $-\frac{1}{2}$.

With the above prior specification, \citet{wangSun} proved that the Bayes factor of Garcia-Donato and Sun (Equation \ref{eq:gds}) simplifies to an analytic expression without integral representation:%
\begin{equation}\label{eq:ws}
 \BF_{10} = \frac{\Gamma\Bigl(\frac{p}{2}+\alpha + \frac{1}{2}\Bigr)\cdot \Gamma\Bigl(\frac{n-p}{2}\Bigr)}{\Gamma\Bigl(\frac{n-1}{2}\Bigr)\cdot \Gamma (\alpha+1)} \cdot \Biggl(\frac{SSR}{SST}\Biggr)^{\alpha-\frac{n-p-2}{2}},
\end{equation}%
With this equation, the Bayes factor can be computed exactly without any need for numerical integration routines. Note: though the gamma function is itself defined by an integral, its exact values can often be found, particularly for positive integers (i.e., $\Gamma(n)=(n-1)!$ for $n\in \mathbb{N}$). In other cases, extremely precise approximations can be had with most standard programming libraries).

\begin{example}\label{example2}
  Let us now compute a Bayes factor using the \citet{wangSun} approach of Equation \ref{eq:ws}. We will consider the same data set given in Example \ref{example1}, so we have $n=18$ and $p=3$. Equation \ref{eq:ws} requires the user to specify $\alpha \in [-\frac{1}{2},0]$, which governs the shape of the Pearson Type VI prior on $\tau$, the ratio of variance components $\sigma^2_a/\sigma^2$. To see the effect of this choice on the resulting Bayes factor, we will consider two cases on the boundary: $\alpha=0$ and $\alpha=-\frac{1}{2}$.\\

  {\it Case 1:} Set $\alpha=0$. Then by Equation \ref{eq:ws}%
  \begin{align*}
    \BF_{10} &= \frac{\Gamma(\frac{3}{2}+0+\frac{1}{2})\cdot \Gamma(\frac{18-3}{2})}{\Gamma(\frac{18-1}{2})\cdot \Gamma(0+1)}\Biggl(\frac{88}{172}\Biggr)^{0-\frac{18-3-2}{2}}\\
            &= \frac{\Gamma(2)\cdot \Gamma(\frac{15}{2})}{\Gamma(\frac{17}{2})\cdot \Gamma(1)} \Biggl(\frac{88}{172}\Biggr)^{-\frac{13}{2}}\\
            &= \frac{1\cdot 1871.254}{14034.41\cdot 1}\Biggr(\frac{88}{172}\Biggr)^{-\frac{13}{2}}\\
    &= 10.393.\\
  \end{align*}

  {\it Case 2:} Set $\alpha=-\frac{1}{2}$. Then by Equation \ref{eq:ws}%
 \begin{align*}
    \BF_{10} &= \frac{\Gamma(\frac{3}{2}-\frac{1}{2}+\frac{1}{2})\cdot \Gamma(\frac{18-3}{2})}{\Gamma(\frac{18-1}{2})\cdot \Gamma(-\frac{1}{2}+1)}\Biggl(\frac{88}{172}\Biggr)^{-\frac{1}{2}-\frac{18-3-2}{2}}\\
            &= \frac{\Gamma(\frac{3}{2})\cdot \Gamma(\frac{15}{2})}{\Gamma(\frac{17}{2})\cdot \Gamma(\frac{1}{2})} \Biggl(\frac{88}{172}\Biggr)^{-7}\\
            &= \frac{0.8862269\cdot 1871.254}{14034.41\cdot 1.772454}\Biggr(\frac{88}{172}\Biggr)^{-7}\\
    &= 7.265.\\
 \end{align*}

 Notice that for both values of $\alpha$, the obtained Bayes factor is quite different from the value obtained from the BIC approximation in Example \ref{example1}. However, both are well under the Sellke et al. bound of 11.10 (which we calculated earlier). 
\end{example}

\section{Analytic Bayes factor from summary statistics}

We are now ready to prove the primary new result in this paper; namely, the development of the Pearson Bayes factor, which is summarized in Proposition \ref{mainTheorem}. The Pearson Bayes factor provides researchers with an analytic Bayes factor (i.e., without integral representation) that can be obtained directly from the summary statistics typically reported from a $t$-test or analysis of variance. To do this, we simply need to start with Equation \ref{eq:ws} of \citet{wangSun}. As we previously saw, their Bayes factor requires knowing the residual sum of squares term $SSR$, the total sum of squares $SST$, the total number of observations $n$, and the number of treatment groups $p$. Additionally, the user must specify a hyperparameter $\alpha$, which controls the shape of the prior distribution for $\tau=\sigma^2_a/\sigma^2$, the ratio of variance components.

To begin the proof, let us first consider two facts. First, $x$ is the between-treatments degrees of freedom, so $x=p-1$. This implies $p=x+1$. Second, $y$ is the residual (within-treatments) degrees of freedom, so $y=n-p$. Putting these two facts together, we have $x+y=(p-1) + (n-p) = n-1$. Substituting these into Equation \ref{eq:ws} gives us%
\begin{align*}\label{eq:proof}
\BF_{10} &= \frac{\Gamma\Bigl(\frac{p}{2}+\alpha + \frac{1}{2}\Bigr)\cdot \Gamma\Bigl(\frac{n-p}{2}\Bigr)}{\Gamma\Bigl(\frac{n-1}{2}\Bigr)\cdot \Gamma (\alpha+1)} \cdot \Biggl(\frac{SSR}{SST}\Biggr)^{\alpha-\frac{n-p-2}{2}}\\
         &= \frac{\Gamma\Bigl(\frac{x+1}{2}+\alpha + \frac{1}{2}\Bigr)\cdot \Gamma\Bigl(\frac{y}{2}\Bigr)}{\Gamma\Bigl(\frac{x+y}{2}\Bigr)\cdot \Gamma (\alpha+1)} \cdot \Biggl(\frac{SSR}{SST}\Biggr)^{\alpha-\frac{y-2}{2}}\\
  &= \frac{\Gamma\Bigl(\frac{x}{2}+\alpha + 1\Bigr)\cdot \Gamma\Bigl(\frac{y}{2}\Bigr)}{\Gamma\Bigl(\frac{x+y}{2}\Bigr)\cdot \Gamma (\alpha+1)} \cdot \Biggl(\frac{SSR}{SST}\Biggr)^{\alpha-\frac{y}{2}+1}.
\end{align*}

The last step is to express the fraction $SSR/SST$ in terms of $F$, $x$, and $y$. To this end, we note that the reciprocal can be written as%
\[
  \frac{SST}{SSR} = \frac{SSR+SSA}{SSR} = 1+\frac{SSA}{SSR}.
\]%
By definition, the $F$ ratio is written as $F=\frac{SSA/x}{SSR/y} = \frac{SSA}{SSR}\cdot \frac{y}{x}$. Combining these gives us%
\[
  \frac{SST}{SSR} = 1+\frac{xF}{y} = \frac{y+xF}{y}.
\]%
Reciprocating and substituting back into the equation above puts the Bayes factor into the form of Proposition \ref{mainTheorem}, thus completing the proof.

\begin{example}\label{example3}
  Again, we consider the same data set as Example \ref{example1}, where we found $F(2,15)=7.16$. From these summary statistics, we apply Proposition \ref{mainTheorem} using $F=7.16$, $x=2$, and $y=15$. As before, we consider the two boundary cases for the value of $\alpha$:\\

  {\it Case 1:} $\alpha=0$. By Proposition \ref{mainTheorem}, we have%
  \begin{align*}
    \PBF_{10} &= \frac{\Gamma\Bigl(\frac{2}{2}+0+1\Bigr)\cdot \Gamma\Bigl(\frac{15}{2}\Bigr)}{\Gamma\Bigl(\frac{2+15}{2}\Bigr)\cdot \Gamma(0+1)}\Biggl(\frac{15}{15+2\cdot 7.16}\Biggr)^{0-\frac{15}{2}+1}\\
            &=\frac{1\cdot 1871.254}{14034.41\cdot 1}(0.5116)^{-6.5}\\
    &= 10.397\\
  \end{align*}

  {\it Case 2:} $\alpha=-1/2$. Here, we have%
  \begin{align*}
    \PBF_{10} &= \frac{\Gamma\Bigl(\frac{2}{2}-\frac{1}{2}+1\Bigr)\cdot \Gamma\Bigl(\frac{15}{2}\Bigr)}{\Gamma\Bigl(\frac{2+15}{2}\Bigr)\cdot \Gamma(-\frac{1}{2}+1)}\Biggl(\frac{15}{15+2\cdot 7.16}\Biggr)^{-\frac{1}{2}-\frac{15}{2}+1}\\
            &=\frac{0.8862269\cdot 1871.254}{14034.41\cdot 1.772454}(0.5116)^{-7}\\
    &= 7.268.\\
  \end{align*}

  In both cases, Proposition \ref{mainTheorem} produces Bayes factors reasonably close to the answers we produced using the Wang and Sun formula (Equation \ref{eq:ws}) in Example \ref{example2}. Indeed, the inaccuracy results from the inevitable two-digit rounding that we often see with summary statistics when reported in papers. If instead we had used $F=7.159$ instead of $F=7.16$, our answers would have matched exactly.

\end{example}

Finally, let us consider the case where $p=2$; that is, we have two treatment conditions. In this case, the between-treatments degrees of freedom is equal to 1. Thus, if we apply Proposition \ref{mainTheorem} with $x=1$, we have the following:%
\begin{align*}
  \PBF_{10} &= \frac{\Gamma\Bigl(\frac{x}{2}+\alpha+1\Bigr)\cdot \Gamma\Bigl(\frac{y}{2}\Bigr)}{\Gamma\Bigl(\frac{x+y}{2}\Bigr)\cdot \Gamma(\alpha+1)}\Biggl(\frac{y}{y+xF}\Biggr)^{\alpha-\frac{y}{2}+1}\\
            &= \frac{\Gamma\Bigl(\frac{1}{2}+\alpha+1\Bigr)\cdot \Gamma\Bigl(\frac{y}{2}\Bigr)}{\Gamma\Bigl(\frac{1+y}{2}\Bigr)\cdot \Gamma(\alpha+1)}\Biggl(\frac{y}{y+F}\Biggr)^{\alpha-\frac{y}{2}+1}\\
  &= \frac{\Gamma\Bigl(\alpha + \frac{3}{2}\Bigr)\cdot \Gamma\Bigl(\frac{y}{2}\Bigr)}{\Gamma\Bigl(\frac{1+y}{2}\Bigr)\cdot \Gamma(\alpha+1)}\Biggl(\frac{y}{y+F}\Biggr)^{(2\alpha-y+2)/2}.
\end{align*}%
If we substitute the identity $F=t^2$ and replace $y$ with $\nu$ (the typical symbol used to represent degrees of freedom in a $t$-test), we get the following:
\begin{align*}
  \PBF_{10} &= \frac{\Gamma\Bigl(\frac{\nu}{2}\Bigr) \cdot \Gamma\Bigl(\alpha + \frac{3}{2}\Bigr)}{\Gamma\Bigl(\frac{\nu+1}{2}\Bigr)\cdot \Gamma(\alpha+1)}\Biggl(\frac{\nu}{\nu+t^2}\Biggr)^{(2\alpha-\nu+2)/2}\\
            &= \frac{\Gamma\Bigl(\frac{\nu}{2}\Bigr) \cdot \Gamma\Bigl(\alpha + \frac{3}{2}\Bigr)}{\Gamma\Bigl(\frac{\nu+1}{2}\Bigr)\cdot \Gamma(\alpha+1)}\Biggl(\frac{\nu+t^2}{\nu}\Biggr)^{(-2\alpha+\nu-2)/2}.\\
            &= \frac{\Gamma\Bigl(\frac{\nu}{2}\Bigr) \cdot \Gamma\Bigl(\alpha + \frac{3}{2}\Bigr)}{\Gamma\Bigl(\frac{\nu+1}{2}\Bigr)\cdot \Gamma(\alpha+1)}\Biggl(1+\frac{t^2}{\nu}\Biggr)^{(\nu-2\alpha-2)/2}
\end{align*}%
which exactly matches the Pearson Bayes factor derived by \citet{wang2016} for the $t$-test. Thus, our proposed Pearson Bayes factor formula from Proposition \ref{mainTheorem} serves as a generalization that works for summary statistics from both the $t$-test and analysis of variance.

\section{Simulation study}

In this final section, I report the results of a simulation study designed to benchmark the performance of the Pearson Bayes factor against other popular Bayes factors. In the simulation, I generated random datasets that each reflected the balanced one-factor designs that we have discussed throughout this paper. Specifically, data were generated as%
\[
  Y_{ij} = \mu + a_j + \varepsilon_{ij};\hspace{5mm}i=1,\dots,r;\hspace{3mm}j=1,\dots,p,
\]%
where $\mu$ represents a grand mean, $a_j \sim \mathcal{N}(0,\sigma_{a})$ represent each of the $p$ randomly drawn treatment effects, and $\varepsilon_{ij} \sim \mathcal{N}(0,\sigma^2)$ represent the normally-distributed error terms. For convenience I set $p=3$, though similar results were obtained with other values of $p$ (not reported here). Also, without loss of generality I set $\mu=0$ and $\sigma=1$. I then systematically varied the following three components of the model:

\begin{enumerate}
\item The number of replicates per treatment condition $r$ was set to either $n=10$, $n=30$, or $n=80$;
\item The size of the overall treatment effect was manipulated by setting $\tau = \sigma_a^2/\sigma^2$ to be either $\tau=0$, $\tau=0.5$, or $\tau=1$. Note that data generated under the condition $\tau=0$ reflect the null model $\h_0$, whereas data generated under $\tau>0$ reflect the alternative model $\h_1$.
\item The prior setting $\alpha$ used in the computation of the Pearson Bayes factor was set at both ends of its consistency range; that is, $\alpha=-1/2$ and $\alpha=0$.
\end{enumerate}

For each combination of number of replicates ($r=10,30,80$), treatment effect ($\tau=0,0.5,1.0$), and prior setting $\alpha$ ($\alpha = -0.5, 0$), I generated 1000 simulated datasets. For each of the datasets, I computed three different Bayes factors: the Pearson Bayes factor proposed in this paper, the JZS Bayes Factor \citep{rouder2012}, and the BIC Bayes factor \citep{faulkenberry2018}.  These Bayes factors were then converted to posterior probabilities via Equation \ref{eq:probability}. To compare the performance of the three Bayes factors in the simulation, I considered three analyses for each simulated dataset: (1) a visualization of the distribution of posterior probabilities $P(\h_0\mid \D)$; (2) a calculation of the proportion of simulated trials for which the correct model was chosen (i.e., model choice accuracy); and (3) a calculation of the proportion of simulated trials for which the Pearson Bayes factor chose the same model as the JZS Bayes factor (i.e., model choice consistency).

\begin{figure}
  \centering
  \includegraphics[width=0.9\textwidth]{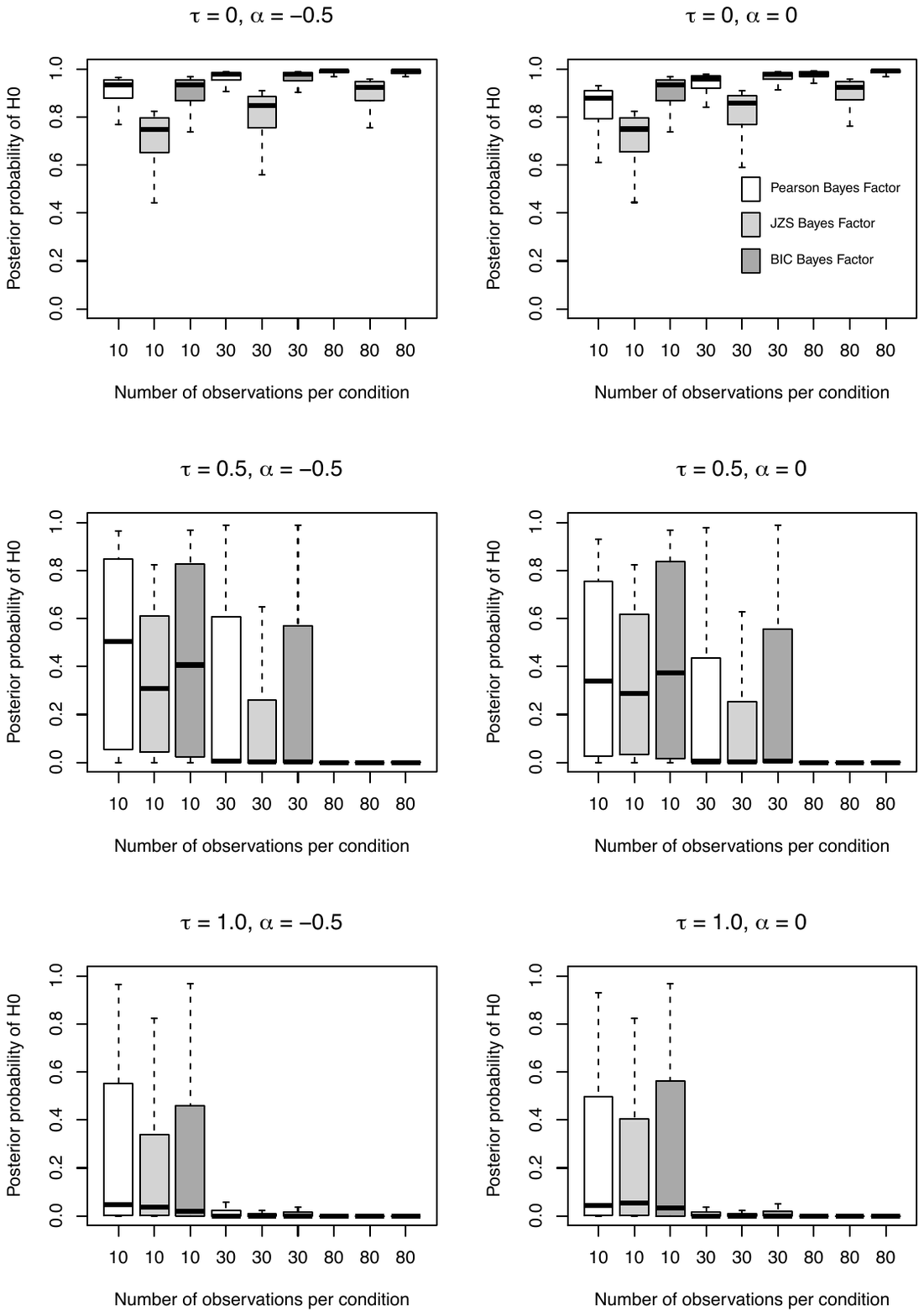}
  \caption{Results from our simulation. Each boxplot depicts the distribution of the posterior probability $P(\h_0\mid \D)$ for 1000 Monte Carlo simulations. White boxes represent posterior probabilities derived from the Pearson Bayes factor. Medium gray boxes represent posterior probabilities derived from the JZS Bayes factor. Dark gray boxes represent posterior probabilities derived from the BIC Bayes factor.}
  \label{fig:dist}
\end{figure}

To begin, let us consider the distribution of posterior probabilities $P(\h_0\mid \D)$. Boxplots of these posterior probabilities can be seen in Figure \ref{fig:dist}. From the figure, we can see that the Pearson Bayes factor performs well compared to the JZS and BIC Bayes factors, but there are some distinct differences. First, when $\tau=0$ (and thus, $\h_0$ is the correct model) both the Pearson and BIC Bayes factors produce larger posterior probabilities for $\h_0$ than the JZS Bayes factor. When the prior parameter $\alpha$ is set to $-1/2$, the Pearson Bayes factor produce a smaller range of posterior probabilities than the BIC Bayes factor. When $\tau>0$ ($\h_1$ is the correct model), the overall advantage goes to the JZS Bayes factor, though the distribution of posterior probabilities becomes indistinguishable across methods as $\tau$ increases and $r$ increases. Note that in the $\tau=0.5$ case (where treatment effect variance is 1/2 of residual variance), model selection is a bit more scattered. In this case, the Pearson Bayes factor outperforms (i.e., smaller range and smaller posterior probability for $\h_0$) the BIC Bayes factor when the prior parameter $\alpha$ is set to 0. This is not surprising, as the Pearson Type VI prior for $\tau$ is more dispersed when $\alpha=0$, allowing for a greater prior mass being placed on larger effects. 

Though the distributions of posterior probabilities tend to follow similar patterns, it is not clear to what extent the three methods provide the user with an accurate inference. To compare the accuracy of the three methods, I calculated {\it model choice accuracy}, defined as the proportion of simulated datasets for which the correct model was chosen. Model choice was defined by considering $\h_0$ to be chosen whenever $\BF_{01}>1$ and $\h_1$ to be chosen whenever $\BF_{01}<1$. The results are displayed in Table \ref{tab:acc}. From the table, we can see confirmation of the intuition gained from the distribution of posterior probabilities in Figure \ref{fig:dist}. Again, for null effects, both the PBF and the BIC consistently outperform the JZS Bayes factor (particularly when the prior shape parameter is set to $\alpha=-1/2$). When $\tau>0$, the the JZS Bayes factor chooses the correct model ($\h_1$) more often than either the PBF or the BIC. Again, the PBF outperforms BIC when setting $\alpha=0$. Accuracy of both PBF and BIC increase as the number of replicates per treatment condition increases.

\begin{table}
  \centering \small
  \begin{tabular}{ccccccccc}
    & & \multicolumn{3}{c}{$\alpha=-1/2$} & & \multicolumn{3}{c}{$\alpha=0$}\\
    \cline{3-5} \cline{7-9}\\
    & & PBF & JZS & BIC && PBF & JZS & BIC\\
    \hline
    $\tau=0$:\\
    $r=10$ & & .980 & .913 & .965 & & .931 & .886 & .978\\
    $r=30$ & & .983 & .933 & .982 & & .978 & .930 & .982\\
    $r=80$ & & .995 & .973 & .995 & & .987 & .966 & .994\\[2mm]

    $\tau=0.5$:\\
    $r=10$ & & .497 & .655 & .551 & & .584 & .652 & .555\\
    $r=30$ & & .732 & .824 & .742 & & .771 & .820 & .733\\
    $r=80$ & & .872 & .914 & .872 & & .892 & .913 & .874\\[2mm]

    $\tau=1.0$:\\
    $r=10$ & & .732 & .830 & .765 & & .751 & .790 & .727\\
    $r=30$ & & .855 & .908 & .859 & & .868 & .900 & .856\\
    $r=80$ & & .932 & .956 & .933 & & .945 & .959 & .938\\
    \hline
    
  \end{tabular}
  \caption{Model choice accuracies for the Pearson Bayes factor (PBF), JZS Bayes factor (JZS), and BIC Bayes factor (BIC), calculated as the proportion of simulated datasets for which the correct model was chosen.}
  \label{tab:acc}
\end{table}

Finally, I assessed the degree to which the model chosen by the Pearson Bayes factor matches the model chosen by the JZS Bayes factor. This model choice consistency is displayed in Table \ref{tab:con}. Overall, model choice consistency is very good (all above 84\%). Certainly, consistency improves as the number of replicates per treatment condition increases. However, it is also notable that consistency improves across the board when $\alpha$ is set to 0. 

\begin{table}
  \centering \small

  \begin{tabular}{ccccc}
    & & $\tau=0$ & $\tau=0.5$ & $\tau=1.0$\\
    \hline
    $\alpha=-1/2$\\
    $r=10$ & & .933 & .842 & .902\\
    $r=30$ & & .950 & .908 & .947\\
    $r=80$ & & .978 & .958 & .976\\[2mm]

    $\alpha=0$\\
    $r=10$ & & .955 & .932 & .961\\
    $r=30$ & & .952 & .951 & .968\\
    $r=80$ & & .979 & .979 & .986\\
    \hline

  \end{tabular}
  \caption{Model choice consistency for the PBF Bayes factor and the JZS Bayes factor, calculated as the proportion of simulated datasets for which both methods chose the same model.}
  \label{tab:con}
\end{table}

\section{Conclusion}
In this paper, I have developed an analytic Bayes factor (the Pearson Bayes factor) which allows a researcher to obtain Bayes factors using the minimal summary statistics (e.g., the test statistic and the degrees of freedom) from both $t$-tests and analysis of variance designs. This formula improves upon the BIC Bayes factor formula of \citet{faulkenberry2018} by not only providing the user with an {\it exact} Bayes factor instead of an approximation, but by also giving the user the ability to specify prior knowledge about the size of effects expected. Our simulation study shows that the Pearson Bayes factor performs comparably to the JZS Bayes factor of \citet{rouder2012} (especially when the prior parameter $\alpha$ is set to 0). Also, the Pearson Bayes factor outperforms JZS on data generated under a null model. The Pearson Bayes factor did not perform as well as the JZS Bayes on data generated under the assumption of nonzero treatment effects, but this limitation is outweighed by the Pearson Bayes factor's unique ability to be computed from summary statistics alone and without the need for computing a multi-dimensional integral. For this reason, the Pearson Bayes factor is an excellent tool for researchers who wish to assess the evidential value of data from published studies in which raw data is not readily available. In all, the Pearson Bayes factor should be a valuable tool in the scientist's statistical toolbox.

\begin{BLReferences}

\bibitem[Faulkenberry, 2018]{faulkenberry2018}
Faulkenberry, T.~J. (2018).
\newblock Computing {B}ayes factors to measure evidence from experiments: {A}n
  extension of the {BIC} approximation.
\newblock {\em Biometrical Letters}, 55(1):31--43.

\bibitem[Faulkenberry, 2019a]{faulkenberry2019}
Faulkenberry, T.~J. (2019a).
\newblock Estimating {B}ayes factors from minimal {ANOVA} summaries for
  repeated-measures designs.
\newblock arXiv:1905.05569.

\bibitem[Faulkenberry, 2019b]{faulkenberry2019ampps}
Faulkenberry, T.~J. (2019b).
\newblock Estimating evidential value from analysis of variance summaries: A
  comment on {Ly} (2018).
\newblock {\em Advances in Methods and Practices in Psychological Science},
  2(4):406--409.

\bibitem[Fisher, 1925]{fisher1925}
Fisher, R.~A. (1925).
\newblock {\em Statistical Methods for Research Workers}.
\newblock Oliver \& Boyd, Edinburgh.

\bibitem[García-Donato and Sun, 2007]{gds}
García-Donato, G. and Sun, D. (2007).
\newblock Objective priors for hypothesis testing in one-way random effects
  models.
\newblock {\em Canadian Journal of Statistics}, 35(2):303--320.

\bibitem[G\"onen et~al., 2005]{gonen2005}
G\"onen, M., Johnson, W.~O., Lu, Y., and Westfall, P.~H. (2005).
\newblock The {B}ayesian two-sample $t$ test.
\newblock {\em The American Statistician}, 59(3):252--257.

\bibitem[Jeffreys, 1961]{jeffreys1961}
Jeffreys, H. (1961).
\newblock {\em The {T}heory of {P}robability (3rd ed.)}.
\newblock Oxford University Press, Oxford, UK.

\bibitem[Kass and Raftery, 1995]{kass1995}
Kass, R.~E. and Raftery, A.~E. (1995).
\newblock Bayes factors.
\newblock {\em Journal of the American Statistical Association}, 90(430):773.

\bibitem[Liang et~al., 2008]{liang2008}
Liang, F., Paulo, R., Molina, G., Clyde, M.~A., and Berger, J.~O. (2008).
\newblock Mixtures of $g$ priors for {B}ayesian variable selection.
\newblock {\em Journal of the American Statistical Association},
  103(481):410--423.

\bibitem[Maruyama, 2009]{maruyama}
Maruyama, Y. (2009).
\newblock {A Bayes factor with reasonable model selection consistency for ANOVA
  model}.
\newblock arXiv:0906.4329v2.

\bibitem[Masson, 2011]{masson2011}
Masson, M. E.~J. (2011).
\newblock A tutorial on a practical {B}ayesian alternative to null-hypothesis
  significance testing.
\newblock {\em Behavior {R}esearch {M}ethods}, 43(3):679--690.

\bibitem[Raftery, 1995]{raftery1995}
Raftery, A.~E. (1995).
\newblock Bayesian model selection in social research.
\newblock {\em Sociological {M}ethodology}, 25:111--163.

\bibitem[Rouder et~al., 2016]{rouder2016}
Rouder, J.~N., Engelhardt, C.~R., McCabe, S., and Morey, R.~D. (2016).
\newblock Model comparison in {ANOVA}.
\newblock {\em Psychonomic Bulletin \& Review}, 23(6):1779--1786.

\bibitem[Rouder et~al., 2012]{rouder2012}
Rouder, J.~N., Morey, R.~D., Speckman, P.~L., and Province, J.~M. (2012).
\newblock Default {B}ayes factors for {ANOVA} designs.
\newblock {\em Journal of {M}athematical {P}sychology}, 56(5):356--374.

\bibitem[Schwarz, 1978]{schwarz1978}
Schwarz, G. (1978).
\newblock Estimating the dimension of a model.
\newblock {\em The Annals of Statistics}, 6(2):461--464.

\bibitem[Sellke et~al., 2001]{sellke2001}
Sellke, T., Bayarri, M.~J., and Berger, J.~O. (2001).
\newblock Calibration of $p$-values for testing precise null hypotheses.
\newblock {\em The American Statistician}, 55(1):62--71.

\bibitem[Student, 1908]{student1908}
Student (1908).
\newblock The probable error of a mean.
\newblock {\em Biometrika}, 6(1):1.

\bibitem[Thiele et~al., 2017]{thiele2017}
Thiele, J.~E., Haaf, J.~M., and Rouder, J.~N. (2017).
\newblock Is there variation across individuals in processing? {B}ayesian
  analysis for systems factorial technology.
\newblock {\em Journal of Mathematical Psychology}, 81:40--54.

\bibitem[Wagenmakers, 2007]{wagenmakers2007}
Wagenmakers, E.-J. (2007).
\newblock A practical solution to the pervasive problems of $p$ values.
\newblock {\em Psychonomic {B}ulletin {\&} {R}eview}, 14(5):779--804.

\bibitem[Wang and Liu, 2016]{wang2016}
Wang, M. and Liu, G. (2016).
\newblock A simple two-sample {B}ayesian $t$-test for hypothesis testing.
\newblock {\em The American Statistician}, 70(2):195--201.

\bibitem[Wang and Sun, 2014]{wangSun}
Wang, M. and Sun, X. (2014).
\newblock Bayes factor consistency for one-way random effects model.
\newblock {\em Communications in Statistics - Theory and Methods},
  43(23):5072--5090.

\bibitem[Wasserstein and Lazar, 2016]{asa}
Wasserstein, R.~L. and Lazar, N.~A. (2016).
\newblock The {ASA} statement on $p$-values: {C}ontext, process, and purpose.
\newblock {\em The American Statistician}, 70(2):129--133.

\bibitem[Zellner, 1986]{zellner1986}
Zellner, A. (1986).
\newblock On assessing prior distributions and {B}ayesian regression analysis
  with $g$-prior distributions.
\newblock In Goel, P.~K. and Zellner, A., editors, {\em Bayesian inference and
  decision techniques: {E}ssays in Honor of {B}runo de {F}inetti}, pages
  233--243. Elsevier.

\end{BLReferences}
\end{document}